\documentclass[twocolumn]{aastex62}

\newcommand{\revision}[1]{\textcolor{black}{#1}}

\graphicspath{{./}{figures/}}
\usepackage{lineno}

\begin{document}

\title{Disentangling the Hadronic Components in NGC 1068}

\author[0000-0002-6584-1703]{Marco Ajello}
\affiliation{Department of Physics and Astronomy, 
Clemson University,
Clemson, SC, 29631}

\author[0000-0002-5358-5642]{Kohta Murase}
\affiliation{
Department of Physics; Department of Astronomy \& Astrophysics; Center for Multimessenger Astrophysics, Institute for Gravitation and the Cosmos, The Pennsylvania State University, University Park, PA 16802, USA
}
\affiliation{School of Natural Sciences, Institute for Advanced Study, Princeton, NJ 08540, USA}
\affiliation{Center for Gravitational Physics and Quantum Information, Yukawa Institute for Theoretical Physics, Kyoto University, Kyoto, Kyoto 606-8502, Japan}

\author[0000-0002-8436-1254]{Alex McDaniel}
\affiliation{Department of Physics and Astronomy, 
Clemson University,
Clemson, SC, 29631}

\begin{abstract}
The recent detection of high-energy neutrinos by IceCube in the direction of the nearby Seyfert/starburst galaxy NGC 1068  implies that radio-quiet active galactic nuclei can accelerate cosmic-ray ions. Dedicated multi-messenger analyses suggest that the interaction of these high-energy ions { with ambient gas or photons} happens in a region of the galaxy that is highly opaque for GeV-TeV gamma rays. Otherwise, the GeV-TeV emission would violate existing constraints provided by {\it Fermi}-LAT and MAGIC. The conditions of high optical depth are realized near the central super-massive black hole (SMBH). At the same time, the GeV emission detected by the {\it Fermi}-Large Area Telescope (LAT) is likely related to the galaxy's sustained star-formation activity. In this work, we derive a 20\,MeV - 1\,TeV spectrum of NGC 1068 using 14\,years of {\it Fermi}-LAT observations. We find that the starburst hadronic component is responsible for NGC 1068's emission above $\sim$500\,MeV. However, below this energy an additional component is required. In the 20-500\,MeV range the {\it Fermi}-LAT data are consistent with  hadronic emission {initiated by non-thermal ions interacting with gas or photons} in the vicinity of the central SMBH. This highlights the importance of the MeV band to discover hidden cosmic-ray accelerators.


\end{abstract}


\section{Introduction} \label{sec:intro}
NGC 1068 is one of the brightest and most studied active galactic nuclei (AGN). It also hosts intense star formation (a starburst). Located at a distance of $\sim$10\,Mpc \citep{curtois2013}, it is classified as a Seyfert 2 galaxy because of the absence of broad emission lines in its optical spectrum \citep{shields75}. However, broad lines have been detected in polarized light \citep{antonucci85}. {This observation represents} one of the foundations of the AGN unification model \citep{urry95} as it implies the presence of an obscuring medium (the torus) on parsec scales.

NGC 1068's prominent X-ray emission is well understood as due to photons from the accretion disk being upscattered to X-rays by a population of thermal electrons located above the accretion disk \citep[the so-called corona,][]{galeev1979,takahara79,haardt1991}. This coronal emission is heavily reprocessed by the dense obscuring torus which is observed nearly edge on \citep[e.g.,][]{bauer2015}.

Recent  observations \citep{garcia2016} with the Atacama Large Millimiter Array (ALMA) have resolved the torus, which is found to have a radius of 3.5\,pc and a mass of $\sim10^5$\,M$_{\odot}$. Further ALMA observations have shown that a wide-angle AGN wind is currently interacting with a large fraction of the molecular torus \citep{garcia2019}.
CO observations indicate the presence of an AGN-driven massive ($>10^7$\,M$_{\odot}$) molecular outflow launched from the inner $\sim100$\,pc region, and a starburst ring located at 1.5\,kpc responsible for most of the galaxy's star-formation rate of $\approx20$\,M$_{\odot}$ yr$^{-1}$ \citep{fluetsch2019}.

The GeV gamma-ray emission of NGC 1068 has typically been ascribed to the star-formation activity, which, through the creation of supernova remnants and pulsar wind nebulae, is able to accelerate cosmic rays \citep{ackermann2012,ajello2020}. Recently,  neutrino emission from NGC 1068 has  been reported by IceCube at a confidence level of 4.2\,$\sigma$ in the 1-20\,TeV energy range \citep{ic_ngc1068}.
In the same energy range, the Major Atmospheric Gamma Imaging Cherenkov  \citep[MAGIC,][]{lorenz2004} telescope reported only upper limits on TeV gamma-ray emission from NGC 1068. 
These upper limits demonstrate that the TeV gamma-ray flux of NGC 1068  is {less than a tenth of the neutrino flux.}
{This implies that the region of hadronic ($pp$ or $p\gamma$) interactions producing the observed neutrinos should be highly opaque to GeV-TeV gamma rays  because hadronic interactions inevitably produce neutrinos and gamma rays with similar energies. Multi-messenger data suggest that the neutrino emission radius $R$ is smaller than $\sim30-100$ Schwartzschild radius \citep[$R_S\sim 0.2$\,AU for NGC 1068,][]{murase2022}.}
Such hidden sources have independently been predicted by the analyses of the all-sky neutrino flux and the diffuse isotropic gamma-ray background \citep{murase2016a,bechtol2017}.   

The conditions of high $\gamma\gamma\rightarrow e^+e^-$ optical depth ($\tau_{\gamma\gamma}$) are reached in the immediate vicinity of the central super-massive black hole (SMBH) \citep[e.g.,][]{murase2016a,murase2020,inoue2020}. 
The GeV-TeV photons are then reprocessed to MeV energies through pair cascades and then leave the source with a spectrum which depends on the distance from the SMBH.
For this reason, in this work, we extract a 20\,MeV-1\,TeV spectrum of NGC 1068 using 14.3\,yr of {\it Fermi} Large Area Telescope \citep[LAT,][]{atwood2009} and interpret it as the sum of two components: a low-energy cascade emission and a high-energy starburst emission. This paper is organized as follows: $\S$~\ref{sec:analysis} describes the analysis of {\it Fermi}-LAT data, $\S$~\ref{sec:model} describes the modeling, while
$\S$~\ref{sec:summary} summarizes the results.
In this work we adopt the standard cosmological parameters: H$_0$=70\,km s$^{-1}$ Mpc $^{-1}$, $\Omega_M=1-\Omega_{\Lambda}$=0.3.

\section{Gamma-Ray Data Analysis}
\label{sec:analysis}
\begin{figure}[htbp]
\includegraphics[width=\linewidth]{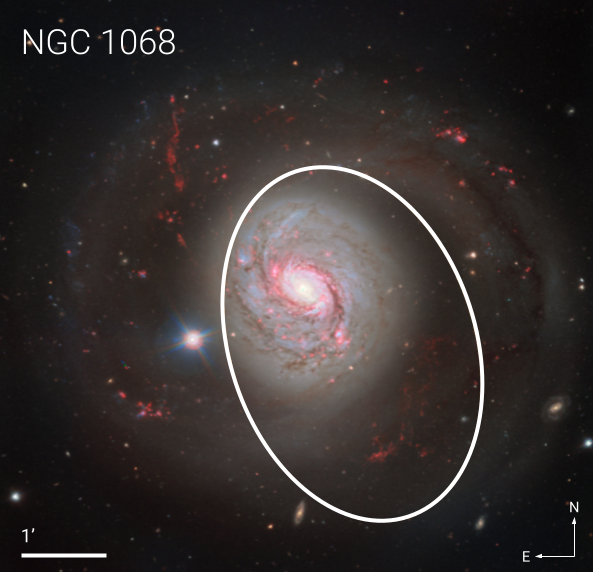}
\caption{95\% positional uncertainty ellipse for NGC 1068 in the 50 MeV - 1 TeV energy range as derived in this analysis overlaid on an image from the VLT. 
\label{fig:NGC_image}
}
\end{figure}

\begin{figure}[htbp]
\includegraphics[scale=1.1]{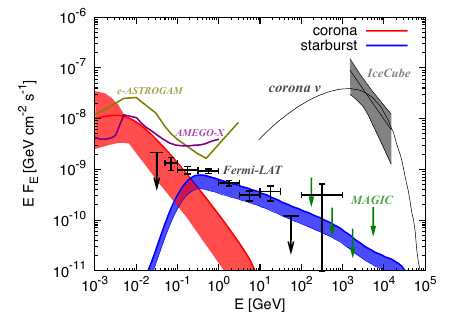}
\caption{Model spectra of MeV-TeV gamma-ray emission from NGC 1068, compared to {\it Fermi}-LAT data obtained by this work (black data points). 
AGN corona \citep{murase2020,murase2022} and starburst \citep[][and see also text]{ajello2020} models are shown by red and blue shaded bands, respectively. 
The all-flavor coronal neutrino spectrum,
which can account for the IceCube data (gray shaded band) \citep{ic_ngc1068}, is also shown with the black thin solid curve \citep{murase2020}. 
Sensitivity curves of {\it AMEGO-X}~\citep{caputo2022} and {\it e-ASTROGAM}~\citep{angelis2017} are also overlaid. 
\label{fig:mmplot}
}
\end{figure}

The gamma-ray data used in this analysis were collected over $\sim 14.3$ years by the {\it Fermi}-LAT between August 4, 2008 and December 1, 2022. The full analysis includes events with energies in the range 20 MeV-1 TeV. We define a $10^{\circ}\times 10^{\circ}$ region of interest (ROI) centered at the 4FGL coordinates of NGC 1068 (4FGL J0242.6-0000). We use the standard data filters (DATA QUAL$>0$ and LAT CONFIG==1) and select photons corresponding to the P8R3\_SOURCE\_V3 class \citep{atwood2013,bruel2018}. The analysis is performed using \texttt{Fermipy} (v1.2, \citealp{fermipy}), which utilizes the underlying Fermitools (v2.2.0). The Galactic diffuse emission is modeled using the standard interstellar emission model (\texttt{gll\_iem\_v07.fits}) and the point source emission is modeled using the 4FGL-DR3 catalog (\texttt{gll\_psc\_v28.fits}, \citealp{4fgl, 4fgl_dr3}). In order to account for photon leakage from sources outside of the ROI due to the PSF of the detector, the model includes all 4FGL sources within a $15^{\circ}\times 15^{\circ}$ region. The energy dispersion correction (edisp\_bins=-1) is enabled for all sources except the isotropic component. The analysis is split between two energy regimes, the 20 MeV - 50 MeV regime and the 50 MeV-1 TeV regime. At the 50 MeV-1 TeV regime we perform a joint likelihood analysis over the four point spread function (PSF) classes  and use a maximum zenith angle of $90^{\circ}$. Each PSF type has a designated isotropic spectrum (iso\_P8R3\_SOURCE\_V3\_PSF{\it i}\_v1, for i
ranging from 0-3) that is used in the analysis. 
{For the 20-50 MeV regime, we adopt the more stringent zenith angle of $80^{\circ}$ and do not differentiate among the different PSF classes. We model the extragalactic emission and residual instrumental background using 
\texttt{iso\_P8R3\_SOURCE\_V3\_v1.txt}. }
{The diffuse emission models are available from the {\it Fermi} Science Support
Center\footnote{\url{https://fermi.gsfc.nasa.gov/ssc/}}}. During the analysis, NGC 1068 is modeled as a power-law source with free index and normalization. The spectral parameters of the Galactic diffuse component (index and normalization) and the normalization of the isotropic component are left free to vary as are the normalizations of all 4FGL sources with TS $\geq$ 25 that are within $5^\circ$ of the ROI center and all sources with TS $\geq$ 500 and within $7^\circ$. \revision{The computation of the spectral energy distribution (SED) data points is performed using the \texttt{sed()} method provided in \texttt{Fermipy}. The spectrum in each energy bin is modeled assuming a power-law with an index of 2 while allowing the normalization of the source to vary.} Upper limits are reported at the 95\% confidence level and are calculated using the Bayesian method \citep{helene1983}. The spectral data from the {\it Fermi}-LAT analysis are listed in Table \ref{tab:sed}.
In Figure \ref{fig:NGC_image} we show an image of NGC 1068 from the European Southern Observatory's (ESO) Very Large Telescope (VLT)\footnote{\url{https://www.eso.org/public/images/eso1720a/}} overlaid with the 95\% positional uncertainty ellipse for the 50 MeV - 1 TeV analysis obtained using the \texttt{localize()} function in \texttt{Fermipy}. {The measured spectrum of NGC 1068, shown in Figure~\ref{fig:mmplot}, is in agreement with the one reported in the 4FGL-DR3 catalog \citep{4fgl_dr3}  and extends it to lower and higher energies, respectively.}
\begin{table}[htbp]
\centering
\renewcommand{\arraystretch}{1.5}
\begin{tabular}{c|c|c}
\hline\hline
$E$ [GeV] & Flux [ergs cm$^{-2}$ s$^{-1}$] & TS\\
\hline
$0.02-0.05$ & $<3.20\times 10^{-12}$ & 0.00\\
$0.05 - 0.10$ & $2.13\pm 0.60\times 10^{-12}$ & 10.23\\ 
$0.10 - 0.32$ & $1.56\pm 0.26\times 10^{-12}$ & 52.76\\ 
$0.32 - 1.00$ & $1.49\pm 0.15\times 10^{-12}$ & 183.48\\ 
$1.00 - 3.16$ & $8.70^{+1.06}_{-1.12}\times 10^{-13}$ & 159.82\\ 
$3.16 - 10.00$ & $4.99^{+1.07}_{-1.21}\times 10^{-13}$ & 56.90\\ 
$10.00 - 31.62$ & $5.86^{+1.67}_{-2.03}\times 10^{-13}$ & 51.11\\ 
$31.62 - 100.00$ & $<1.90\times 10^{-13}$ & 0.00\\ 
$100.00 - 1000.00$ &  $4.95^{+3.36}_{-4.81}\times 10^{-13}$ & 7.86\\
\hline\hline
\end{tabular}\caption{SED values for NGC 1068 between 20 MeV $-$ 1 TeV. Upper limits are reported at the 95\% confidence level and are computed using the Bayesian method.} \label{tab:sed}
\end{table}

\section{Models and Implications}
\label{sec:model}
\subsection{Starburst Galaxies}
NGC 1068 is  one of the starburst galaxies detected by the {\it Fermi}-LAT~\citep{ackermann2012,ajello2020}, and it was also considered to be among the most promising {sources of PeV neutrinos}~\citep{murase2016b,lamastra2016}. The starburst region is considered to be transparent to GeV-TeV gamma rays, and the observed GeV gamma-ray emission presumably comes from the decay of neutral pions, although the neutrino flux {that modeling predicts would be associated with} the gamma-ray emission is too low to explain the IceCube data. 

Assuming that the starburst region is nearly calorimetric \citep[see e.g.,][]{mcdaniel2023},
{we calculate the gamma-ray emission produced by cosmic rays via inelastic $pp$ interactions with interstellar gas},
adopting the method used in \cite{murase2022}. The normalization of the starburst model is set by the $L_\gamma-L_{\rm IR}$ relation obtained by \cite{ajello2020}, where $\log_{10} L_{\rm IR}=10.97$\footnote{This is the luminosity at a distance of 10\,Mpc.} is used for NGC 1068 \citep{sanders2003}. In Figure~\ref{fig:mmplot}, we show the $2\sigma$ uncertainty bands for the starburst model. 
It is known that pionic gamma rays have a spectral break around 0.1~GeV below which the gamma-ray spectrum falls as $E F_E\propto E^2$. Figure~\ref{fig:mmplot} shows an excess of the data over the starburst model, particularly for energies at $\lesssim$500\,MeV. 

We also note that GeV gamma-ray emission could be produced by cosmic rays accelerated by AGN, perhaps through disk winds~\citep{liu2017,ajello2021}.  
Indeed, the source luminosity as predicted by the $L_\gamma-L_{\rm IR}$ relation slightly underestimates the 
true  luminosity measured by {\it Fermi}-LAT \citep[see also][]{yoasthull2014}.  
\cite{inoue2022} proposed that the observed GeV gamma-ray emission may originate from interactions between the disk wind and the dusty torus. However, the sub-GeV excess exists even for these scenarios as long as the primary gamma-ray {emission is produced primarily by hadronuclear interactions.} {Finally, in starburst galaxies the leptonic component is sub-dominant to the hadronic one and its spectrum is harder than the excess observed here \citep{yoasthull2014,peretti2019}.}

\subsection{AGN Coronae}
The excess of $\lesssim$500\,MeV gamma-ray emission shown in Figure~\ref{fig:mmplot} suggests the presence of another component at these energies. It may be a hint of gamma-ray emission from the coronal regions around the AGN accretion disk. It is widely believed that a hot, strongly-magnetized plasma, the so-called ``corona'', may produce   X-ray emission through Compton upscattering of disk photons \citep{galeev1979,haardt1991}. Recent global magnetohydrodynamic simulations~\citep[e.g.,][]{jiang2019} and particle-in-cell simulations~\citep[e.g.,][]{groselj2023} have demonstrated that such magnetically-powered coronal regions naturally form as a result of magnetic dissipation in the black hole accretion system. 

\cite{murase2020} proposed the magnetically-powered corona model for multi-TeV neutrino emission in which cosmic rays are accelerated by magnetic dissipation and the resulting turbulence in the vicinity of SMBHs. High-energy protons interact with optical/UV photons from the accretion disk and X-rays from the corona via $p\gamma$ interactions as well as the coronal gas via $pp$ interactions. {They showed the importance of  Bethe-Heitler pair production for the energy losses of the protons making TeV neutrinos, as well as  calculated the cascade emission resulting from synchrotron, inverse Compton, and two-photon annihilation.} The model not only explains  the multi-TeV neutrino flux of NGC 1068 but also the all-sky neutrino intensity in the 10~TeV range; furthermore, it predicts the associated proton-induced cascade gamma-ray emission in the MeV range. The cascade emission largely originates from synchrotron emission for strongly magnetized coronae.  

In Figure~\ref{fig:mmplot}, the cascade gamma-ray spectrum of the magnetically-powered corona model is taken from \cite{murase2020}, where an emission radius of $R=30R_S$  and an intrinsic X-ray luminosity of $L_X=(1-3)\times10^{43}~{\rm erg}~{\rm s}^{-1}$ are used\footnote{The cosmic-ray pressure $P_{\rm CR}$ is set to $15-50$\% of the virial pressure, adopting a distance of $10$~Mpc.}. The corresponding neutrino spectrum explains the observed IceCube data for NGC 1068 (see Figure~\ref{fig:mmplot}). Interestingly, the cascade gamma-ray emission accompanied by neutrinos may explain the sub-GeV excess indicated by our {\it Fermi}-LAT analysis. 

The break or cutoff energy of the coronal gamma-ray emission, which is set by $\tau_{\gamma\gamma}\sim1$, depends on $R$ and $L_X$. While predictions for hadronic gamma-ray emission at $\sim1-10$~MeV energies are rather robust, the flux in the $\sim0.1$~GeV range can be lower for smaller values of $R$ \citep{murase2022}. For this reason, in Figure~\ref{fig:mmplot}, {the red uncertainty band of the model} includes the case of $R=3R_s$ (corresponding to the innermost stable circular orbit radius of a non-rotating black hole) and $L_X=7\times10^{43}~{\rm erg}~{\rm s}^{-1}$ (corresponding to the maximum luminosity within the $1\sigma$ uncertainty of {\it NuSTAR} observations) and considers both the minimal $pp$ and $p\gamma$ models in \cite{murase2022}.

\section{Summary and Discussion}
\label{sec:summary}
In this work, we have measured the gamma-ray spectrum of NGC 1068 using 14.3\,yr of {\it Fermi}-LAT observations. We have, for the first time, extended the measurement to 20\,MeV to constrain potential hadronic components whose gamma-ray emission is absorbed and reprocessed in the MeV band \citep{murase2020,inoue2022}. 
We have found that above $\gtrsim$500\,MeV, the NGC 1068 spectrum can be well explained as the product of star-formation activity. This emission is mostly hadronic in origin, particularly in starburst galaxies like NGC 1068, which act  nearly as proton calorimeters \citep{lacki2011}. Indeed, in these galaxies the primary and secondary leptonic components are sub-dominant to the $\pi^0$-decay component \citep{yoasthull2014,peretti2019}, although we do not exclude 
a potential contribution to the gamma-ray flux
from electrons accelerated by outflows~\citep{lenain2010,lamastra2016}. 
\revision{The low-energy ($<$500\,MeV) part of the spectrum can be described by the proton-induced cascade emission from $pp$ interaction and/or Bethe-Heitler pair-production processes in the AGN corona. This hadronic component is able to explain the low-energy {\it Fermi} and IceCube data at the same time, as shown in Figure~\ref{fig:mmplot}. This implies that the two hadronic components that contribute to the NGC 1068 spectrum arise from different regions in the host galaxies.}

\begin{figure}[ht!]
\includegraphics[width=\linewidth]{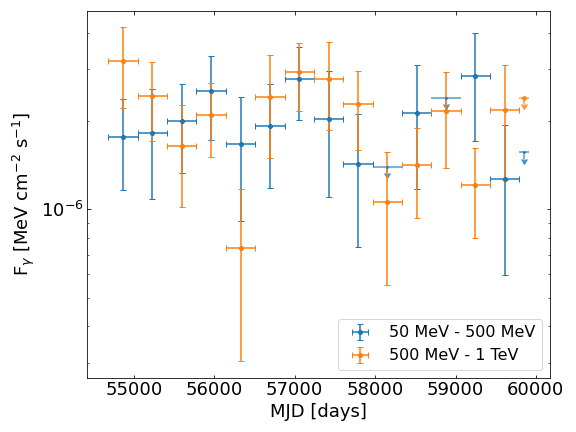}\\
\caption{Yearly lightcurve of NGC 1068 at low (50 - 500 MeV, blue) and high (500 MeV - 1 TeV, orange) energies. 
\label{fig:lc_plot}
}
\end{figure}

In the magnetically-powered corona model \citep{murase2020}, it is natural to expect time variability for coronal neutrino and gamma-ray emission. The minimum variability timescale can be the light-crossing time, $R/c$, which may range from minutes to hours. Longer variability timescales of days or longer -- associated with dissipation, rotation and accretion -- are also possible. {To test this scenario, we extracted a low-energy (50-500\,MeV) and a high-energy (500\,MeV-1\,TeV) yearly (because of the low signal-to-noise ratio) binned lightcurve of the source. These lightcurves are reported in Figure~\ref{fig:lc_plot} and show no evidence of variability (p-values of 0.40 and 0.15, respectively)\footnote{The significance of variability has been computed as in Appendix A.3 of \cite{ajello2020}.}.}

Furthermore, we note that the highest energy photon detected by the {\it Fermi}-LAT within 0.25\,deg of NGC 1068 (well within the 95\,\% containment radius at $>$100\,GeV) has an energy of 738\,GeV. The second most energetic photon has an energy of  217\,GeV. This shows that future observations by the Cherenkov Telescope array \citep[CTA,][]{CTA} may detect the high-energy emission of NGC 1068.

MeV gamma-ray emission may also be produced by non-thermal electrons~\citep{inoue2020}. Particle acceleration mechanisms are currently uncertain, and not only stochastic acceleration in turbulence~\citep{murase2020} but also magnetic reconnection~\citep{kheirandish2021} and shock acceleration~\citep{stecker1991,inoue2020,inoue2022} have been proposed. 
Further multi-messenger and multi-wavelength studies, including MeV gamma-ray observations with e.g., {\it AMEGO-X}, \citep{caputo2022}, will enable us to probe the physics of dissipation and particle acceleration in the coronal regions.


\begin{acknowledgements}
MA and AM acknowledge support from NASA grant 80NSSC22K1580. Clemson University is acknowledged for generous allotment of compute time on Palmetto cluster.
KM is partly supported by the NSF Grants No.~AST-1908689, No.~AST-2108466 and No.~AST-2108467, and KAKENHI No.~20H01901 and No.~20H05852.

The \textit{Fermi} LAT Collaboration acknowledges generous ongoing support
from a number of agencies and institutes that have supported both the
development and the operation of the LAT as well as scientific data analysis.
These include the National Aeronautics and Space Administration and the
Department of Energy in the United States, the Commissariat \`a l'Energie Atomique
and the Centre National de la Recherche Scientifique / Institut National de Physique
Nucl\'eaire et de Physique des Particules in France, the Agenzia Spaziale Italiana
and the Istituto Nazionale di Fisica Nucleare in Italy, the Ministry of Education,
Culture, Sports, Science and Technology (MEXT), High Energy Accelerator Research
Organization (KEK) and Japan Aerospace Exploration Agency (JAXA) in Japan, and
the K.~A.~Wallenberg Foundation, the Swedish Research Council and the
Swedish National Space Board in Sweden.
 
Additional support for science analysis during the operations phase is gratefully
acknowledged from the Istituto Nazionale di Astrofisica in Italy and the Centre
National d'\'Etudes Spatiales in France. This work performed in part under DOE
Contract DE-AC02-76SF00515.
\end{acknowledgements}



\end{document}